\documentclass{elsart}
\usepackage{amssymb}
\usepackage{graphicx}

\input{tcilatex}

\begin{document}

\begin{frontmatter}
\title{Highly $^{13}$C isotope enriched azafullerene, C$_{59}$N, for nuclear spin labelling} 
\author{$^{1}$ F. Simon \corauthref{email}},
\author{$^{2}$F. F\"{u}l\"{o}p},
\author{$^{2,3}$A. Rockenbauer},
\author{$^{2,3}$L. Korecz},
\author{$^{1}$H. Kuzmany},

\address{$^{1}$ Institut f\"{u}r Materialphysik, Universit\"{a}t Wien, Strudlhofgasse 4, A-1090 Wien, Austria}
\address{$^{2}$ Budapest University of Technology and Economics, Institute of Physics and Solids in Magnetic
Fields Research Group of the Hungarian Academy of Sciences, H-1521, Budapest P.O.Box 91, Hungary}
\address{$^{3}$ Chemical Research Center of the Hungarian Academy of Sciences, P.O.Box 17, Budapest, H-1525 Hungary}

\begin{abstract}
Synthesis of highly $^{13}$C isotope enriched azafullerene, C$_{59}$N embedded in C$_{60}$ is reported. $^{13}$C
enriched fullerenes, produced with the Kr\"{a}tschmer-Huffmann process, were subject to a N$_{2}$ discharge that
produces C$_{59}$N with a low probability. Raman spectroscopy indicates a homogeneous 
$^{13}$C distribution. Electron spin resonance measurement (ESR) proves that the C$_{59}$N 
concentration, 0.2 \%, is similar as in non-enriched fullerenes. The ESR 
spectrum is simulated accurately with the known $^{14}$N and $^{13}$C hyperfine coupling constants. 
The material enables the nuclear spin-labelling of heterofullerene complexes 
with a potential for biological applications. It might also find applications as a building element for quantum computation.
\end{abstract}

\corauth[email]{Corresponding author, email: fsimon@ap.univie.ac.at}%


\end{frontmatter}


\section{Introduction}

Isotope controlled synthesis (ICS) of molecular nanostructures provides an
important degree of freedom to characterize fundamental and application
oriented properties. ICS is generally considered as a tool to e.g. enhance
nuclear magnetic resonance signals, to provide improved information when
specific isotope labelling is possible or to trace biological processes
using radioactive nuclei. For fullerenes \cite{Kroto}, ICS was applied to
improve the NMR data \cite{PenningtonRMP}, to identify the origin of
different vibrational modes in crystalline C$_{60}$ \cite{MihalyPRB1995},
and to yield an insight into underlying physical phenomenon such as the
mechanism of the superconductivity in alkali doped fullerides \cite{GunnRMP}
by means of $^{13}$C enrichment.\ More recently, $^{13}$C enriched
fullerenes were used to produce $^{13}$C enriched single wall carbon
nanotubes \cite{CM0406343}.

Properties of fullerenes can be also studied through the synthesis of
on-ball doped modifications. The C$_{59}$N or C$_{59}$B heterofullerenes
were predicted to provide a doping opportunity for C$_{60}$ \cite%
{AndreoniCPL1992}\cite{WudlReview}. In general, heterofullerenes possess a
rich chemistry due to their enhanced reactivity as compared to pristine
fullerenes \cite{WudlReview}. The C$_{59}$N azafullerene can be synthesized
in macroscopic amounts \cite{WudlReview}\cite{WudlSCI} and in a solid form
it is an insulator consisting of (C$_{59}$N)$_{2}$ dimer units where the
extra electrons are localized in the dimer bonds as singlet states \cite%
{PichlerPRL1997}. The C$_{59}$N monomer radical can be observed by light 
\cite{DinseJACS}\cite{HirschJACS} or thermal induced homolysis of (C$_{59}$N)%
$_{2}$ \cite{SimonJCP} or when the C$_{59}$N monomer is embedded in a low
concentration in the C$_{60}$ crystal \cite{FulopCPL}. This C$_{59}$N:C$%
_{60} $ solid solution was synthesized in a discharge tube designed for the
production of N@C$_{60}$ \cite{FulopCPL}. The advantages of the latter
synthesis method over the chemical synthesis \cite{WudlSCI} are its relative
simplicity and the ability of providing an isotope control option by
changing the $^{14}$N$_{2}$ gas to $^{15}$N$_{2}$. Recently, it was shown
that the extra electron on the C$_{59}$N is transferred toward the C$_{60}$%
's at high temperatures and it provides a controllable electron doping of
the crystalline C$_{60}$ \cite{Rockenbauerunpub}.

Here, we report a combination of the two synthesis routes:\ the $^{13}$C
isotope controlled synthesis of the C$_{59}$N monomer radical. The material
was prepared from C$_{60}$ containing isotopically controlled amounts of $%
^{13}$C using the N$_{2}$ discharge method. Raman spectroscopy indicates a
uniform $^{13}$C enrichment of the fullerenes. The $^{13}$C enriched C$_{59}$%
N:C$_{60}$ material was studied with electron spin resonance. The $^{14}$N
hyperfine triplet, that dominates the spectrum for non isotope enriched C$%
_{59}$N:C$_{60}$, collapses into a broad line in agreement with the isotope
content and the $^{13}$C nuclear hyperfine couplings determined previously
in C$_{59}$N \cite{FulopCPL}. A minority phase that is poor in $^{13}$C was
also observed underlying the sensitivity of the ESR method in characterizing
this material.

\section{Experimental}

\textit{Sample preparation.} Commercial $^{13}$C isotope enriched fullerene
mixture (MER Corp., Tucson, USA) was used for the synthesis of C$_{59}$N.
The isotope enriched fullerenes were produced by the Kr\"{a}tschmer-Huffmann
process \cite{HuffmannNAT} using $^{13}$C enriched graphite rods. The
supplier provided a $^{13}$C enrichment of nominal 25 \% that was determined
using mass spectrometry. The isotope enriched fullerenes are denoted as $%
\left( ^{13}\text{C}_{x}\right) _{60}$ and $\left( ^{13}\text{C}_{x}\right)
_{59}$N in the following. We refer to the material with the nominal $x=0.25$ 
$^{13}$C content, although this value is slightly refined in this work.
Apart from the C$_{70}$ and other higher fullerenes with contents up to 20
\%, the dominant impurity in the material is a $\left( ^{13}\text{C}%
_{x}\right) _{60}$ phase with $x\approx 0.05$ and a content below 2 \%. The
high purity C$_{60}$ (\TEXTsymbol{>}99.9 \%) used for comparison was
obtained from Hoechst (Hoechst AG, Frankfurt, Germany). C$_{59}$N production
was performed in the same N$_{2}$ discharge tube as previously \cite%
{FulopCPL} following the original design of Pietzak \textit{et al.} \cite%
{PietzakCPL} for the production of N@C$_{60}$:C$_{60}$. In brief, fullerenes
are sublimed into a nitrogen discharge that is maintained by a high voltage
between two electrodes inside a quartz tube filled with a low pressure of N$%
_{2}$ gas. C$_{59}$N:C$_{60}$ deposits on surfaces with temperatures between
200-400 $^{\circ }$C of the quartz tube, whereas N@C$_{60}$:C$_{60}$
deposits on the water-cooled cathode. The resulting material is collected
from the tube walls and is resublimed at 500 $^{\circ }$C twice in order to
remove impurities that are usually produced during the synthesis and to
reduce the amount of higher fullerenes \cite{Dresselhaus}. However, the less 
$^{13}$C enriched C$_{60}$ phase can not be removed from the sample with
this method.  Typically 10 mg of final material containing $\left( ^{13}%
\text{C}_{0.25}\right) _{59}$N at 2000 ppm concentrations in $\left( ^{13}%
\text{C}_{0.25}\right) _{60}$ is produced from 100 mg starting fullerene
material. The samples were sealed in quartz tubes under vacuum for the ESR
and Raman measurements. Among the higher fullerenes, C$_{70}$ can be best
observed using Raman spectroscopy and its amount was found to be below 1 \%
in the final material.

\textit{Raman spectroscopy.} Multi frequency Raman spectroscopy was studied
on a Dilor xy triple spectrometer at 488 nm excitation energy using an Ar-Kr
mixed-gas laser.

\textit{Electron spin resonance spectroscopy.} The ESR experiments were
performed with a Bruker Elexsys X-band spectrometer. A typical microwave
power of 1 mW and 0.01 mT magnetic field modulation at ambient temperature
were used.

\begin{figure}[tbp]
\includegraphics[width=0.8\hsize]{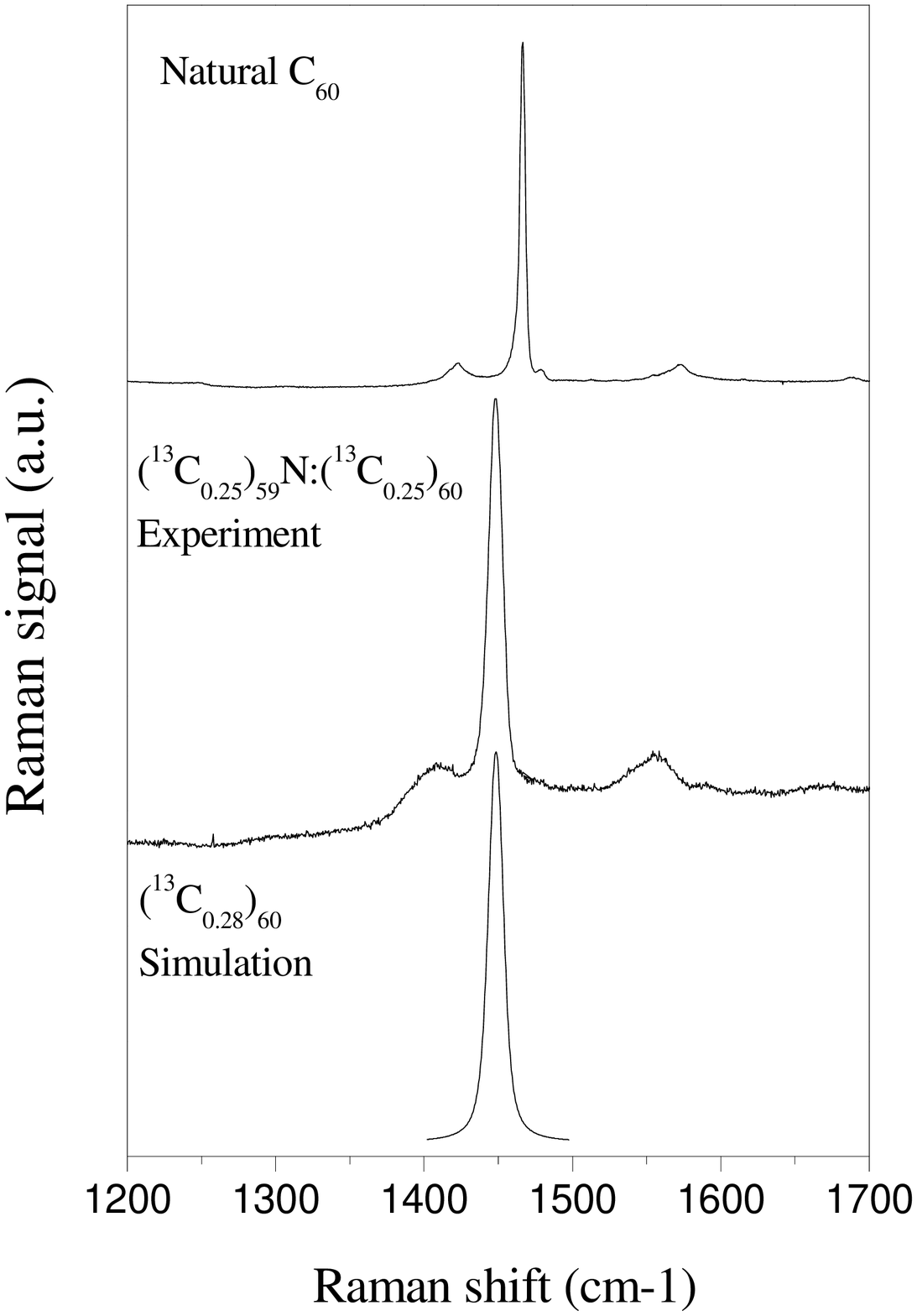}
\caption{Raman spectra of natural C$_{60}$ and $( ^{13}$C$_{0.25}) _{59}$N:$( ^{13}$C$_{0.25}) _{60}$ at $\protect\lambda$ = 488 nm excitation. The lowest solid curve show the
simulated C$_{60}$ A$_{g}$(2) mode with 28 per cent $^{13}$C enrichment as
explained in the text.}
\label{Ramanspectra}
\end{figure}

\section{Results and discussion}

In Fig. \ref{Ramanspectra}., we show the Raman spectra of $\left( ^{13}\text{%
C}_{0.25}\right) _{59}$N:$\left( ^{13}\text{C}_{0.25}\right) _{60}$ and C$%
_{60}$ with natural carbon isotope contents at ambient conditions and
excited with a 488 nm laser. The spectrum of $\left( ^{13}\text{C}%
_{0.25}\right) _{59}$N:$\left( ^{13}\text{C}_{0.25}\right) _{60}$ is
identical to that of $\left( ^{13}\text{C}_{0.25}\right) _{60}$ as the Raman
technique is not sensitive to the 2000 ppm amounts of $\left( ^{13}\text{C}%
_{0.25}\right) _{59}$N. We focus our attention on the totally symmetric A$%
_{g}$(2) mode that appears with the largest intensity for this laser
excitation \cite{Dresselhaus}. Analysis of this mode enables us to determine
the $^{13}$C enrichment level with precision and provides information on its
homogeneity. A similar analysis was performed previously \cite%
{KendzioraPRB1995}\cite{HoroyskiPRB1996}. As the $^{13}$C build-in in the C$%
_{60}$'s is a random process, the number of $^{13}$C nuclei on a given $%
\left( ^{13}\text{C}_{x}\right) _{60}$ fullerene is expected to follow a
binomial distribution with $x\cdot 60$ expectation value. The vibrational
frequency of the $^{13}$C enriched fullerenes downshifts as a result of the
heavier $^{13}$C. In a continuum approximation the amount of the downshift
is given by: $\left( \nu _{0}-\nu \right) /\nu _{0}=1-\sqrt{\frac{12+c_{0}}{%
12+c}}$, where $\nu _{0}$ and $\nu $ are the Raman shifts of the
corresponding modes in the natural carbon and enriched materials,
respectively, $c$ is the concentration of the $^{13}$C enrichment, and $%
c_{0}=0.011$ is the natural abundance of $^{13}$C in carbon. The
experimentally observed 16.0(5) cm$^{-1}$ downshift of the first moment of
the A$_{g}$(2) mode, corresponds to $c=0.28(1)$. In addition, the full
line-shape was simulated from the convolution of the binomial distribution
with the line-shape of this mode in the natural C$_{60}$. A good agreement
between the experimentally observed line-shape and the simulation (lowest
solid curve in Fig. \ref{Ramanspectra}) was obtained using the above value
for $c$. This proves that the distribution of the $^{13}$C nuclei follows
the statistical expectation and is therefore homogeneous. The current $%
c=0.28(1)$ is slightly different from the value, $c=0.25$, given by the
supplier underlining the difficulty of the $^{13}$C content determination.

In order to further characterize the material, we compared its Raman spectra
with that of the starting fullerene mixture (not shown). The absence of the C%
$_{70}$ peaks puts a 2 \% upper limit on the total amount of residual higher
fullerenes in our material as compared to the starting value of about 20 \%.
This proves that the double sublimation procedure at 500 $^{\circ }$C is
indeed very effective in purifying the material from higher fullerenes. The $%
\left( ^{13}\text{C}_{x}\right) _{60}$ phase with $x\approx 0.05$ that was
observed with mass spectroscopy by the supplier is not removed by the double
sublimation, however its content is below the detectability limit. In
contrast, electron spin resonance spectroscopy can detect the fraction of
the sample that belongs to the different $\left( ^{13}\text{C}_{x}\right)
_{59}$N radicals.

\begin{figure}[tbp]
\includegraphics[width=0.8\hsize]{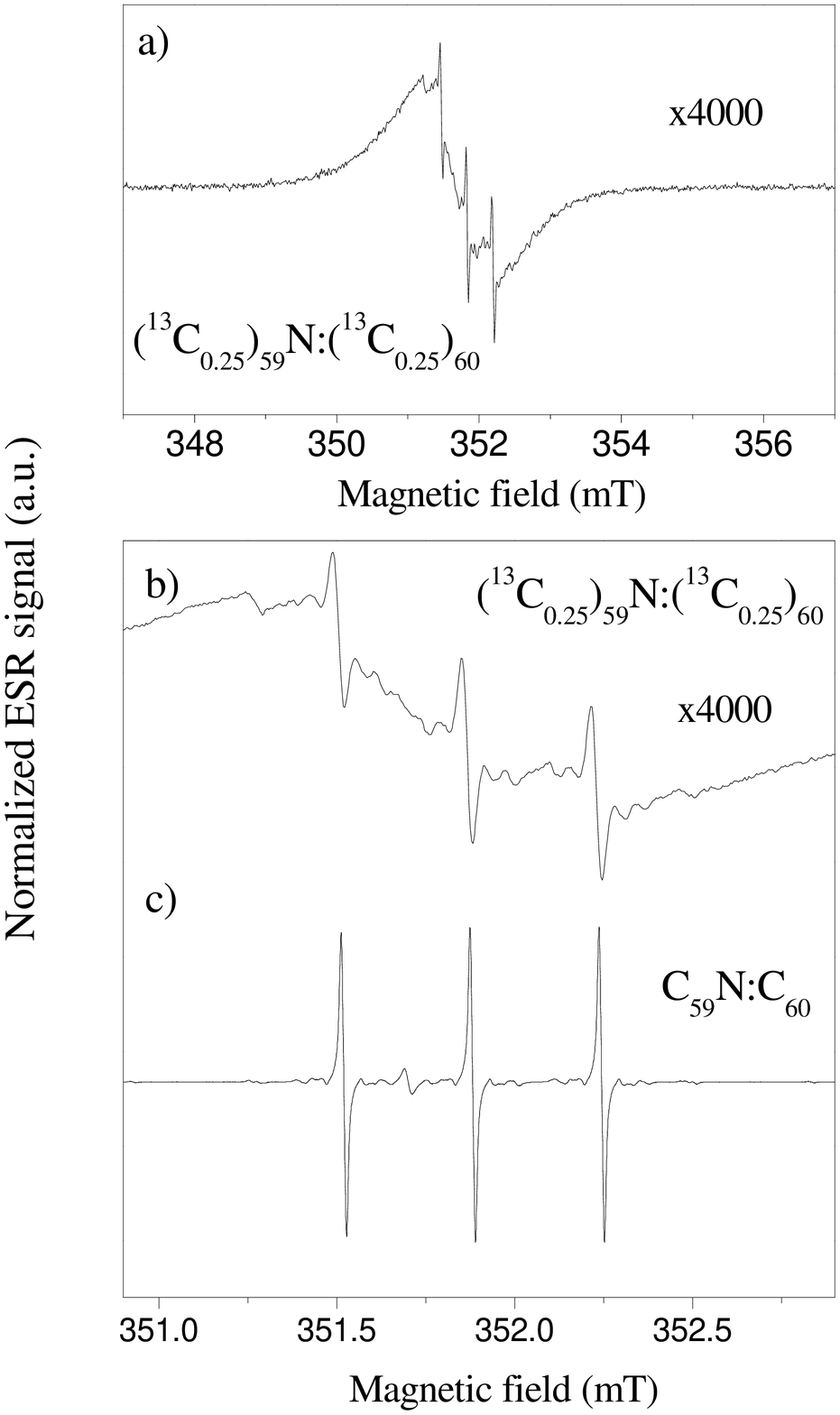}
\caption{ESR spectra of C$_{59}$N:C$_{60}$ produced from a-b) $^{13}$C
enriched and c) natural carbon normalized by the sample mass. Note the
different field scales for a) and b-c) and the enlarged scaling for the
enriched materials.}
\label{ESR1}
\end{figure}

In Fig. \ref{ESR1}., we show the room temperature ESR spectrum of the $%
\left( ^{13}\text{C}_{0.25}\right) _{59}$N$:\left( ^{13}\text{C}%
_{0.25}\right) _{60}$ material with two different magnetic field scales. We
also show the spectrum of non-enriched C$_{59}$N:C$_{60}$ for comparison.
The spectra of the latter was analyzed in detail previously \cite{FulopCPL}
and is recalled here. The dominating triplet component was identified as due
to the $^{14}$N ($I=1$) hyperfine interaction. The free tumbling of the
molecule above the \textit{sc-fcc} structural transition of the C$_{60},$ $%
T_{c}=261$ K \cite{Dresselhaus}, averages out the anisotropic part of this
hyperfine coupling. However, the electron is delocalized on the C$_{59}$N
cage and a number of well defined $^{13}$C ($I=1/2$) satellite doublets
appear as a result of the finite electron density on the different C
positions and the 1.1 \% abundance of $^{13}$C in carbon. The hyperfine
couplings were determined for eight non-equivalent carbon sites
corresponding to twenty-three sites on the C$_{59}$N molecule. An additional
small intensity signal between the two low-field $^{14}$N triplet lines was
identified as a C$_{59}$N$^{+}$-C$_{60}^{-}$ heterodimer due to a partial
charge transfer from C$_{59}$N \cite{Rockenbauerunpub}.

The ($^{13}$C$_{x}$)$_{59}$N spectrum is simulated for arbitrary $x$ by
using a recursive build-up technique \cite{Rocky}. The effect of the first $%
^{13}$C coupling is computed by superimposing a doublet pattern with
intensity $x$ onto each components of the original nitrogen triplet signal
with intensity of $1-x$. In the next step, this superimposed spectrum is
considered as a starting signal and the next superimposition is carried out
in the same way by using the next carbon splitting constant. The new carbon
splitting could have the same value as the preceding one in the case of
equivalent carbons. The procedure is repeated for all the 23 carbon nuclei
with resolved splitting. The impact of the 36 non-resolved carbon splittings
can be considered by using an increased intrinsic line-width in the primer
spectrum. Similarly, the small intensity C$_{59}$N$^{+}$-C$_{60}^{-}$
heterodimer signal becomes unobservable due to its broadening.

The ESR spectrum of the $\left( ^{13}\text{C}_{0.25}\right) _{59}$N:$\left(
^{13}\text{C}_{0.25}\right) _{60}$ consists of apparently two overlapping
signals:\ a broad component and a triplet signal together with a $^{13}$C
hyperfine pattern. The $^{13}$C hyperfine structure is similar to that
observed in the non-enriched C$_{59}$N, however its components have an
increased line-width that results from the hyperfine interaction of the
non-resolved C sites. The double integrated and mass normalized intensities,
that measures the number of spins in the sample, are similar for the natural
and enriched materials suggesting that the complicated pattern in the
enriched sample also originates from the C$_{59}$N radicals, however, the
presence of $^{13}$C broadens its spectrum. Below, we show that the observed
ESR pattern reflects the inhomogeneity in our sample. The narrow triplet and
the broader components originate from the less and highly $^{13}$C enriched
phases, respectively.

\begin{figure}[tbp]
\includegraphics[width=0.8\hsize]{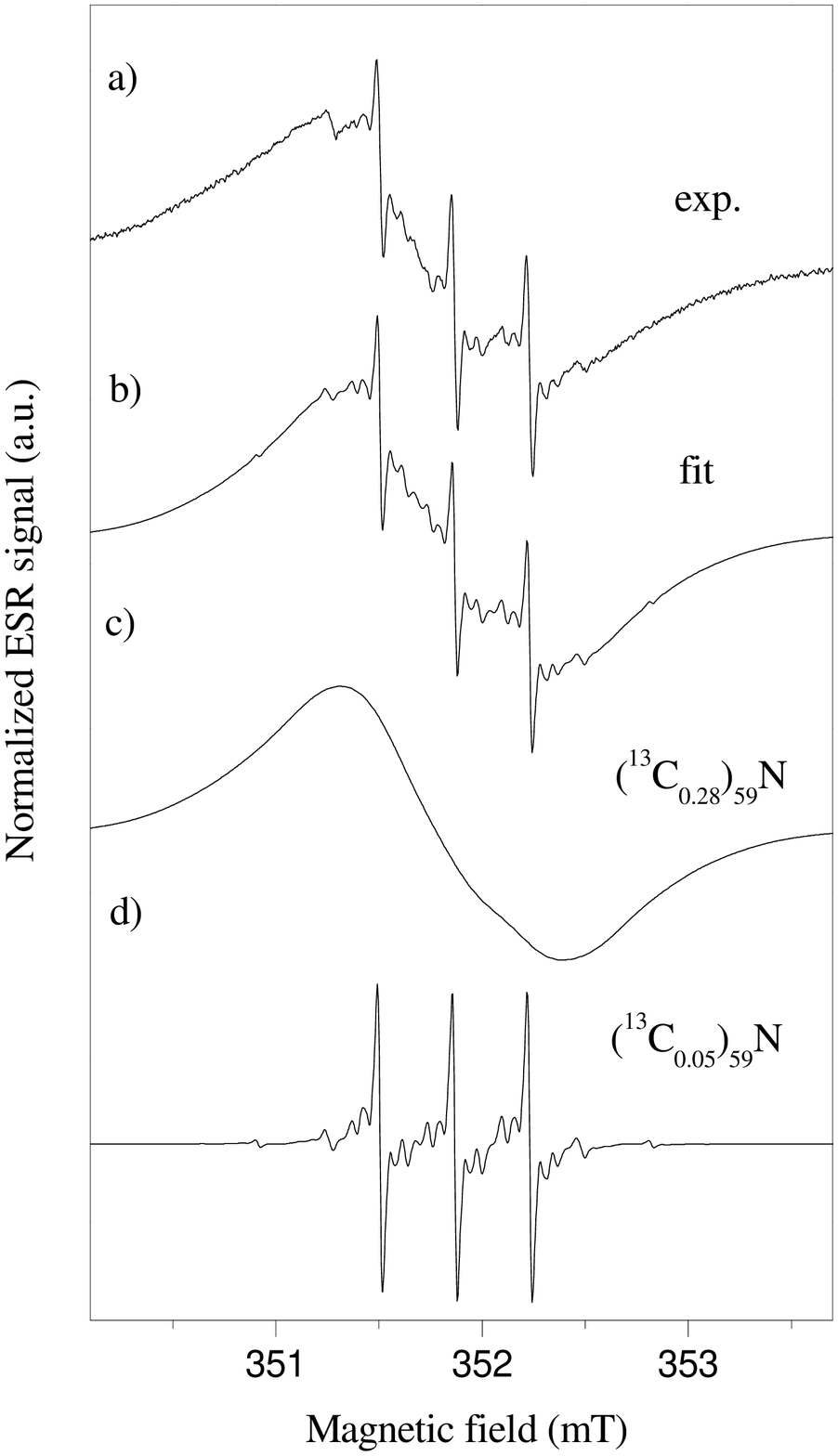}
\caption{Comparison of the experimental ESR spectra of $( ^{13}$C$_{0.25}) _{59}$N:$( ^{13}$C$_{0.25})_{60}$ (a) with
the simulation as explained in the text (b). Simulated spectra for the two
levels of isotope enrichments, $( ^{13}$C$_{0.28}) _{59}$N:$( ^{13}$C$_{0.25}) _{60}$ (c) and $( ^{13}$C$_{0.05}) _{59}$N:$( ^{13}$C$_{0.25}) _{60}$ (d) are
also shown.}
\label{ESRdeconv}
\end{figure}

In Fig. \ref{ESRdeconv}a. we show the experimental spectrum again together
with a simulation for a sample containing a mixture of $\left( ^{13}\text{C}%
_{x}\right) _{59}$N molecules with $x=0.28$ and $x=0.045$ enrichments (Fig. %
\ref{ESRdeconv}b.) with intensity ratios of 98.2:1.8, respectively. The 
\textit{g}-factors of the two components were taken to be identical with
that of the non-enriched C$_{59}$N of $g=2.0014(2)$. A residual 0.04 mT
line-width was assumed to account for the hyperfine interactions of the
non-resolved C sites. The combination of these spectra was found to simulate
best the experimental curve. It was assumed that the molecules are freely
rotating and the spectra can be described using the previously determined
hyperfine coupling constants. The simulated spectra are shown separately for
the two types of molecules in Fig. \ref{ESRdeconv}c. and d., respectively.
Although, the two kinds of molecules have a similar ESR amplitude, they have
very different integrated intensities. We recall that the ESR signal
intensity is inversely proportional to the square of the linewidth due to
the field modulation technique employed. As a result, the narrow structure
is only a tiny, \TEXTsymbol{<}2 \% fraction of the total ESR signal
intensity. This small amount of $^{13}$C poorer phase is an unwanted
side-product of the production of the higher $^{13}$C enriched material,
however, its amount may not be a limiting factor for practical applications.

\section{Conclusion}

In conclusion, we presented the preparation of a $^{13}$C enriched
heterofullerene, the $^{13}$C$_{59}$N azafullerene, from $^{13}$C enriched
fullerenes. Raman and ESR spectroscopy was used to characterize the
enrichment and its homogeneity. The material was produced in a nitrogen
discharge tube with the same yield as the non-enriched material. This
synthesis method opens new prospects for applications of the chemically
active heterofullerenes. These include nuclear spin-labelling of
bio-molecules with heterofullerenes or the nuclear spin labelling of the
biologically active fullerene itself such as in the HIV-1 inhibitor
fullerene derivatives \cite{FriedmanJACS}. In addition, an emerging field
where application of the current system is envisaged is the use of molecules
with well defined interaction configurations between electron and nuclear
spins for the purpose of quantum computing \cite{HarneitPSS}\cite{MehringPRL}%
\cite{MehringKB}.

\section{Acknowledgement}

This work was supported by the Austrian Science Funds (FWF) project Nr.
17365, by the EU project NANOTEMP BIN2-2001-00580, PATONN
MEIF-CT-2003-501099 and the Hungarian State Grants OTKA T043255 and T046953.


\begin{thebibliography}{99}
\bibitem{Kroto} H. W. Kroto, J. R. Heath, S. C. O'Brien, R. F. Curl, and R.
E. Smalley, Nature 318 (1985) 162.

\bibitem{PenningtonRMP} For a review see: C. H. Pennington and V. A.
Stenger, Rev. Mod. Phys. 68 (1996) 885.

\bibitem{MihalyPRB1995} M. C. Martin, J. Fabian, J. Godard, P. Bernier, J.
M. Lambert, and L. Mih\'{a}ly, Phys. Rev. B 51 (1995) 2844.

\bibitem{GunnRMP} For a review see: O. Gunnarsson, Rev. Mod. Phys. 69 (1997)
575 and "Alkali-doped Fullerides. Narrow-band solids with unusual
properties" World Scientific, Singapore, 2004.

\bibitem{CM0406343} F. Simon, Ch. Kramberger, R. Pfeiffer, H. Kuzmany, V.
Zolyomi, J. Kurti, P. M. Singer, H. Alloul, cond-mat/0406343.

\bibitem{AndreoniCPL1992} W. Andreoni, F. Gygi, M. Parrinello, Chem. Phys.
Lett. 190 (1992) 159.

\bibitem{WudlReview} J. C. Hummelen, C. Bellavia-Lund, and F. Wudl:\
Heterofullerenes in Topics in current chemistry, vol. 199, springer, Berlin,
Heidelberg, 1999, p. 93.

\bibitem{WudlSCI} J. C. Hummelen, B. Knight, J. Pavlovich, R. Gonzalez, and
F. Wudl, Science 269 (1995) 1554.

\bibitem{PichlerPRL1997} T. Pichler, M. Knupfer, M. S. Golden, S. Haffner,
R. Friedlein, J. Fink, W. Andreoni, A. Curioni, M. Keshavarz-K, C.
Bellavia-Lund, A. Sastre, J.-C. Hummelen, and F. Wudl, Phys. Rev. Lett. 79
(1997) 3026.

\bibitem{DinseJACS} A. Gruss, K.-P. Dinse, A. Hirsch, B. Nuber, and U.
Reuther, J. Am. Chem. Soc. 119 (1997) 8728.

\bibitem{HirschJACS} K. Hasharoni, c. Bellavia-Lund, M. Keshavarz-K, G.
Srdanov, and F. Wudl, J. Am. Chem. Soc. 119 (1997)\ 11128.

\bibitem{SimonJCP} F. Simon, D. Ar\v{c}on, N. Tagmatarchis, S. Garaj, L. Forr%
\'{o} and K. Prassides, J. Phys. Chem. A. 103 (1999) 6969.

\bibitem{FulopCPL} F. F\"{u}l\"{o}p, A. Rockenbauer, F. Simon, S. Pekker, L.
Korecz, S. Garaj, and A. J\'{a}nossy, Chem. Phys. Lett. 334 (2001) 233.

\bibitem{Rockenbauerunpub} A. Rockenbauer \textit{et al.} unpublished.

\bibitem{HuffmannNAT} W. Kr\"{a}tschmer, L. D. Lamb, K. Fostiropoulos, and
D. R. Huffmann, Nature 347 (1990) 354.

\bibitem{PietzakCPL} B. Pietzak, M. Waiblinger, T.A. Murphy, A. Weidinger,
M. Hohne, E. Dietel, A. Hirsch, Chem. Phys. Lett. 279 (1997) 259.

\bibitem{Dresselhaus} M. S. Dresselhaus, G. Dresselhaus, P. C. Ecklund:
Science of Fullerenes and Carbon Nanotubes, Academic Press, 1996.

\bibitem{KendzioraPRB1995} A. Rosenberg and C. Kendziora, Phys. Rev. B 51
(1995) 9321.

\bibitem{HoroyskiPRB1996} P. J. Horoyski, M. L. W. Thewalt, and T. R.
Anthony, Phys. Rev. B 54 (1996) 920.

\bibitem{Rocky} A. Rockenbauer and L. Korecz, Appl. Magn. Reson. \textbf{10}
(1996) 29.

\bibitem{FriedmanJACS} S. H. Friedman, D. L. DeCamp, R. P. Sijbesma, G.
Srdanov, F. Wudl, and G. L. Kenyon, J. Am. Chem. Soc. 115 (1993) 6506.

\bibitem{HarneitPSS} W. Harneit, C. Meyer, A. Weidinger, D. Suter, J.
Twamley, Phys. Stat. Sol. B 233 (2002) 453.

\bibitem{MehringPRL} M. Mehring, J. Mende, and W. Scherer, Phys. Rev. Lett.
90 (2003) 153001.

\bibitem{MehringKB} W. Scherer, A. Weidinger, and M. Mehring in H. Kuzmany,
J. Fink, M. Mehring, S. Roth (Eds.) Electronic Properties of Synthetic
Nanostructures, AIP Conference Proceedings, New York, 2004, p. 315.
\end{thebibliography}
\end{document}